\documentclass[twocolumn,showpacs,prd,floatfix,axodraw]{revtex4}
\usepackage{mathrsfs}
\usepackage{graphicx,,booktabs,bm}
\usepackage{overpic}
\usepackage{color}
\usepackage{amssymb}

\usepackage{mathrsfs,bm,amsmath,amssymb}
\usepackage{longtable,lscape}
\usepackage{txfonts}
\usepackage{amssymb}
\usepackage{indentfirst}
\usepackage{graphicx,,booktabs}
\usepackage{multirow}
\usepackage{color}
\usepackage{amssymb}

\definecolor{cover}{rgb}{0.77,0.87,0.88}
\definecolor{blueone}{rgb}{0.1,0.1,.7}
\definecolor{citec}{rgb}{0.14,0.47,0.09}
\definecolor{two}{rgb}{0.0,0.5,0.}
\definecolor{three}{rgb}{.5,.1,0.15}
\usepackage[bookmarks=true,bookmarksopen=false,plainpages=false,breaklinks=true,
   bookmarksnumbered=true,hypertexnames=false,
   filecolor=blue,urlcolor=three,menucolor=three,
   linkcolor=three,citecolor=blueone, colorlinks,
   anchorcolor=blue,runcolor=pink,frenchlinks=red
   pdfstartview=FitH,pdftitle=title,%
   pdfauthor=author]{hyperref}

\begin{document}
\title{{Internal structures of the baryons $\Xi(2030)$ and $\Xi(2120)$}}

\author{Hao Hei}
\author{Yin Huang\footnote{corresponding author}}\email{huangy2019@swjtu.edu.cn}
\affiliation{ School of Physical Science and Technology, Southwest Jiaotong University, Chengdu 610031,China}

\date{\today}
\begin{abstract}
Currently, there is much controversy surrounding the interpretation of the $\Xi(2030)$ and $\Xi(2012)$ as traditional hadrons containing double strange quarks.
In particular, the ratios of the partial decay widths into $\Lambda\bar{K}$ and $\Sigma\bar{K}$ for $\Xi(2030)$ cannot obtain a suitable explanation under the
$qss$ three quark structure~\cite{Xiao:2013xi}.  Thus, we suggest the $\Xi(2030)$ and $\Xi(2012)$ to be $VB(=\bar{K}^{*}\Sigma/\rho\Xi/\bar{K}^{*}\Lambda/\phi\Xi/\omega\Xi)$
molecular states.  In this work, we perform a systematical investigation of possible  molecular states from the $VB(=\bar{K}^{*}\Sigma/\rho\Xi/\bar{K}^{*}\Lambda/\phi\Xi/\omega\Xi)$  interaction.  The interaction of the system considered is described by the $t$-channel vector ($\rho,\omega,\phi,\bar{K}^{*}$) and pseudoscalar ($\pi,\eta,\bar{K}$) mesons exchanges.  By solving the non-relativistic Schr\"{o}dinger equation with the obtained one-boson-exchange potentials, the $VB(=\bar{K}^{*}\Sigma/\rho\Xi/\bar{K}^{*}\Lambda/\phi\Xi/\omega\Xi)$ bound states with different quantum numbers are searched.  The calculation suggests that $\Xi(2030)$ can be assigned as a $P$-wave $\bar{K}^{*}\Sigma/\rho\Xi/\bar{K}^{*}\Lambda/\phi\Xi/\omega\Xi$ molecular state with spin parity $J^P=5/2^{+}$. The calculation also predict the existence of four
$\bar{K}^{*}\Sigma/\rho\Xi/\bar{K}^{*}\Lambda/\phi\Xi/\omega\Xi$ bound states with $J^P=1/2^{\pm}$ and $J^P=3/2^{\pm}$.  The $\Xi(2012)$ may be a candidate for one of these
four bound states.  If $\Xi(2012)$ is an $S$-wave molecular state with $J^P=1/2^{-}$ or $J^P=3/2^{-}$, we suggest determining its spin and parity by studying its decay
width, owing to the difference in their molecular components.
\end{abstract}

%\pacs{13.60.Le, 12.39.Mk,13.25.Jx}

\maketitle
\section{INTRODUCTION}
Thanks to significant progress in experiments over the past few decades, researchers have discovered numerous new hadrons~\cite{ParticleDataGroup:2022pth}.
While some of these hadrons exhibit a straightforward quark-antiquark configuration for mesons or a three-quark configuration for
baryons~\cite{Godfrey:1985xj,Capstick:1986ter}, many states possess enigmatic structures that defy conventional quark-state understanding.
These exotic states are often considered to be hadron-hadron molecules, opening the door for the exploration of structural hadrons.

The pursuit of possible hadron-hadron molecular structures represents a pivotal aspect of hadron spectroscopy, offering insights into the mechanisms
governing quark dynamics and baryon formation. The deuteron, a well-known bound state comprising neutron-proton components, serves as an early example.
Assuming that $\Lambda(1405)$ is a bound state of $\bar{K}N$~\cite{Oset:1997it,Oller:2000fj,Nemoto:2003ft,Hall:2014uca}, this interpretation effectively
addresses the mass inversion problem.  Even more promising than the deuteron and $\Lambda(1405)$ is the discovery of the $X(3872)$~\cite{BESIII:2013ris,Belle:2013yex}.
Due to its charge and its proximity to the $\bar{D}^{}D$ threshold in terms of mass, the molecular state interpretation of $\bar{D}^{}D$ was proposed\cite{Wang:2013cya,He:2014nya,Wilbring:2013cha}. In 2019, the LHCb Collaboration reported three narrow hidden-charm pentaquark states, named $P_c(4312)$,
$P_c(4440)$, and $P_c(4457)$\cite{LHCb:2019kea}. Their spectroscopic properties and decay widths find a compelling explanation within the context of
$\Sigma_c\bar{D}^{(*)}$ molecular states\cite{Chen:2019bip,Guo:2019fdo,Xiao:2019aya,He:2019ify,Xiao:2019mvs}.  More recently, the LHCb Collaboration
unveiled another novel hidden-charm pentaquark state, $P_{cs}(4459)$\cite{LHCb:2020jpq}, which can be assigned as a $\Xi_c\bar{D}^{*}$ molecular state\cite{Chen:2020uif,Peng:2020hql,Wang:2020eep,Chen:2020kco,Yang:2021pio}. The existence of additional candidates for molecular states has been
explored in Ref.~\cite{Guo:2017jvc}.

Inspired by the above observation and their interpretation as molecules, it is interesting to study whether there exist hadronic molecular states
corresponding to $\Xi$ baryon.  At present, there are eleven $\Xi$ baryons listed in the review of Particle Data Group (PDG)~\cite{ParticleDataGroup:2022pth}.
The ground-state octet and decuplet baryons, the $\Xi(1320)$ and the $\Xi(1530)$, are well established with four star ratings and can be easily fitted into
the conventional quark model.   For the states $\Xi(1690)$, $\Xi(1820)$, $\Xi(2030)$ and $\Xi(2120)$, there exist many different interpretations, such as
$qqq$ states, molecular systems, etc~\cite{Sekihara:2015qqa,Kolomeitsev:2003kt,Sarkar:2004jh,Chao:1980em,Pervin:2007wa,Xiao:2013xi,Khemchandani:2016ftn}.
In particular, we will show that there is good reason to believe that the state $\Xi(2030)$ is a $P$-wave bound state in this work.

The $\Xi(2030)$ is a three-star state and has a mass of $2023\pm{}5$ MeV and a width of $20^{+15}_{-5}$ MeV~\cite{ParticleDataGroup:2022pth}.  An early experimental
analysis~\cite{Amsterdam-CERN-Nijmegen-Oxford:1977bvi} suggested that the spin of the $\Xi(2030)$ should be $J\geq{}5/2$.  Before the experimental observation
of the $\Xi(2030)$, Samios et al. predicted that according to the SU(3) flavor symmetry the $\Xi(2030)$ is most likely the partner of the $N(1680)$, $\Lambda(1820)$,
and $\Sigma(1915)$ with $J^P=5/2^{+}$~\cite{Samios:1974tw}.  The constituent quark model calculations in Ref.~\cite{Chao:1980em} indicated that the $\Xi(2030)$
might be a candidate for the $J^P=5/2^{+}$ state or $J^P=7/2^{+}$ state.  However, the strongly decay analysis based on the experimental measurements disfavors
the assignment of the $\Xi(2030)$ as a member of the $5/2^{+}$ octet~\cite{PavonValderrama:2011gp}.  Also, the strong decay analysis in the chiral quark model~\cite{Xiao:2013xi}
concludes that the $\Xi(2030)$ could not be assigned as any spin-parity $J^P=7/2^{+}$ states or pure $J^P=5/2^{+}$ state, It seems to favor the $J^P = 3/2^{+}$
assignment.  However, this conflicts with the early analysis of the data~\cite{Amsterdam-CERN-Nijmegen-Oxford:1977bvi}.

Compared with the $\Xi(2030)$, the experimental information on the one-star state $\Xi(2120)$ is scarce~\cite{ParticleDataGroup:2022pth}: both the spin parity and
the width is not known experimentally.  The $\Xi(2120)$ was first observed in the $\bar{K}\Lambda$ invariant mass spectrum by the
Amsterdam-CERN-Nijmegen-OxfordCollaboration~\cite{Amsterdam-CERN-Nijmegen-Oxford:1976ezm}, and later confirmed by the French-Soviet and CERN-Soviet Collaboration
~\cite{French-Soviet:1979pox} in the 1970s, where a mass of about 2120 MeV and a width of about 20 MeV were suggested by those observations with poor statistics.
There exist a few theoretical studies about the nature of the $\Xi(2120)$.  A study in the chiral unitary approach suggested that a pole around 2100 MeV can be produced
from the interaction between pseudoscalar/vector mesons and baryons with $J^P=1/2^{-}$ and $3/2^{-}$, which can be associated to the $\Xi(2120)$~\cite{Oset:2010tof,Gamermann:2011mq}.
In Ref.~\cite{Khemchandani:2016ftn} the $I(J^P)=1/2(3/2^{-})$ state located at $2046-i8.2$ MeV is identified as a meson-baryon molecule that can be associated to the
$\Xi(2120)$.  Meanwhile, it is claimed that the $\Xi(2120)$ have a big $\bar{K}\Sigma$ component.

Based on the preceding discussion, it is evident that the $\Xi(2030)$ cannot be easily explained as a $qqq$ state. The mass of the $\Xi(2030)$ closely approaches the
thresholds for $\bar{K}^{*}\Sigma$ and $\rho\Xi$, suggesting a possible interpretation as a bound state of $\bar{K}^{*}\Sigma/\rho\Xi$.  Furthermore, we observe strong
coupling between the $\bar{K}^{*}\Sigma$ and $\rho\Xi$ channels with other channels like $\bar{K}^{*}\Lambda$, $\omega\Xi$, and $\phi\Xi$~\cite{Khemchandani:2016ftn}.
And the $\Xi(2120)$ has been previously interpreted as a $\bar{K}^{*}\Sigma-\rho\Xi-\bar{K}^{*}\Lambda-\phi\Xi-\omega\Xi$ molecular state in the literature~\cite{Khemchandani:2016ftn,Oset:2010tof,Gamermann:2011mq}. Therefore, it is
plausible to interpret the $\Xi(2030)$ and $\Xi(2120)$ baryons as two distinct bound states originating from the $\bar{K}^{*}\Sigma-\rho\Xi-\bar{K}^{*}\Lambda-\phi\Xi-\omega\Xi$ interaction with differing quantum numbers.  Assuming this interpretation, the $\Xi(2030)$ is, at the very least, a $P$-wave bound state. Extensive evidence, as
presented in references~\cite{Kang:2016zmv,He:2016pfa,Zhu:2021exs,Jian:2022rln}, supports the existence of several $P$-wave bound state candidates with molecular
components such as $B^{(*)}_s\pi-\bar{B}^{(*)}K$, $\bar{D}^{*}\Sigma_c$, $D_s\bar{D}_{s0}(2370)$, and $\bar{B}^{(*)}N$.

In this study, we explore the interactions of $\bar{K}^{*}\Sigma-\rho\Xi-\bar{K}^{*}\Lambda-\phi\Xi-\omega\Xi$ and aim to elucidate the properties of the baryons $\Xi(2030)$ and $\Xi(2012)$.
In addition, our also predicts the existence of other molecular states composed of $\bar{K}^{*}\Sigma-\rho\Xi-\bar{K}^{*}\Lambda-\phi\Xi-\omega\Xi$ components.
This paper is organized as follows. In Sec.~\ref{Sec: formulism}, we will present the theoretical formalism. In Sec.~\ref{Sec: results}, the numerical result will be given,
followed by discussions and conclusions in last section.

\section{THEORETICAL FORMALISM}\label{Sec: formulism}
\begin{figure}[h!]
\centering
\includegraphics[bb=80 590 500 710, clip, scale=0.55]{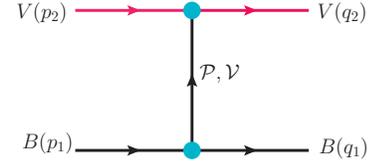}
\caption{The Feynman diagrams for the vector meson-baryon interaction via $t$-channel mesons exchange. }\label{fig.exchange}
\end{figure}

In this work, we search for possible vector meson-baryon molecular states in one-boson exchange (OBE) model, which has been successfully employed to study the $NN$ interaction~\cite{Machleidt:1987hj}.  With great success in reproducing the $NN$ data, the one-boson exchange (OBE) model was also extended to study the
systems with heavy flavors in Refs.~\cite{Chen:2015loa,Chen:2016ypj,Wang:2019nwt,Wang:2020bjt,Li:2012cs,Sun:2011uh}.  Especially, it is a powerful tool in reproducing
the observed pentaquarks~\cite{Chen:2015loa}.  Our calculation is based on a non-relativistic approach in one-boson exchange (OBE) model, which only contain
the one-meson-exchange diagrams.  Here, the channels involved are $\bar{K}^{*}\Sigma,\rho\Xi,\bar{K}^{*}\Lambda,\phi\Xi$, and $\omega\Xi$, and the relevant Feynman diagrams are illustrated in Fig.~\ref{fig.exchange}.

In order to compute the potential kernel, the gauge invariant hidden local symmetry Lagrangian for the coupling of vector mesons to
the baryon octet given by~\cite{Khemchandani:2016ftn}
\begin{align}
{\cal{L}}_{VBB}&=-g\{\langle\bar{B}\gamma_{\mu}[V^{\mu},B]\rangle+\langle\bar{B}\gamma_{\mu}B\rangle\langle{}V^{\mu}\rangle\nonumber\\
               &+\frac{1}{4M}(F\langle\bar{B}\sigma_{\mu\nu}[V^{\mu\nu},B]\rangle+D\langle\bar{B}\sigma_{\mu\nu}\{V^{\mu\nu},B\}\rangle)\},
\end{align}
where the $M$ represent the masses of the baryon and $\langle...\rangle$ refers to an $SU(3)$ trace. $\{A,B\}=AB+BA$ and $[A,B]=AB-BA$.
The constants $D=2.4$ and $F=0.82$ were found to reproduce well the magnetic moments of the baryons~\cite{Jido:2002yz}.  The tensor field
of the vector mesons $V^{\mu\nu}=\partial^{\mu}V^{\nu}-\partial^{\nu}V^{\mu}$ and $\sigma^{\mu\nu}=\frac{i}{2}(\gamma^{\mu}\gamma^{\nu}-\gamma^{\nu}\gamma^{\mu})$.
$B$ and $V^{\mu}$ are the SU(3) baryon octet and vector meson matrices, respectively,
\begin{align}
B&=
\left(
  \begin{array}{ccc}
    \frac{1}{\sqrt{2}}\Sigma^{0}+\frac{1}{\sqrt{6}}\Lambda & \Sigma^{+}                                                &  p                                \\
    \Sigma^{-}                                             & -\frac{1}{\sqrt{2}}\Sigma^{0}+\frac{1}{\sqrt{6}}\Lambda   &  n                                \\
     \Xi^{-}                                               & \Xi^{0}                                                   &  -\frac{2}{\sqrt{6}}\Lambda       \\
  \end{array}
\right),\\
V_{\mu}&=
\left(
  \begin{array}{ccc}
    \frac{1}{\sqrt{2}}(\rho^{0}+\omega) & \rho^{+}                             &  K^{*+}     \\
    \rho^{-}                            & \frac{1}{\sqrt{2}}(-\rho^{0}+\omega) &  K^{*0}     \\
     K^{*-}                             & \bar{K}^{*0}                         &  \phi       \\
  \end{array}
\right)_{\mu}.
\end{align}

The Lagrangians involving the interaction of the three-vector mesons vertex $VVV$ and the coupling of two vector mesons to pseudoscalar meson
vertex $VVP$ are given by~\cite{Oset:2002sh,Gonzalez:2008pv}
\begin{align}
&{\cal{L}}_{VVV}=ig\langle(V^{\mu}\partial_{\nu}V_{\mu}-\partial_{\nu}V_{\mu}V^{\mu})V^{\nu}\rangle,\\
&{\cal{L}}_{VVP}=\frac{G^{'}}{\sqrt{2}}\epsilon^{\mu\nu\alpha\beta}\langle\partial_{\mu}V_{\nu}\partial_{\alpha}V_{\beta}P\rangle,
\end{align}
where $\epsilon^{\mu\nu\alpha\beta}$ is the Levi-Civit\`{a} tensor with $\epsilon^{0123}=1$.  The $P$ is standard SU(3)
pseudoscalar meson octet matrix
\begin{align}
P=
\left(
  \begin{array}{ccc}
    \frac{1}{\sqrt{2}}\pi^{0}+\frac{1}{\sqrt{6}}\eta       & \pi^{+}                                                   &  K^{+}                         \\
    \pi^{-}                                                & -\frac{1}{\sqrt{2}}\pi^{0}+\frac{1}{\sqrt{6}}\eta         &  K^0                           \\
     K^{-}                                                 & \bar{K}^{0}                                               &  -\frac{2}{\sqrt{6}}\eta       \\
  \end{array}
\right).
\end{align}
The coupling constant $G^{'}=\frac{3g^{'}}{4\pi^2f}$ with $g^{'}=-\frac{G_Vm_{\rho}}{\sqrt{2}f^2}$, $G_V=55$ MeV, and $f=93$ MeV.
The coupling constant $g$ can be fixed from the strong decay width of $K^{*}\to{}K\pi$.  With help of the following Lagrangian
\begin{align}
{\cal{L}}_{VPP}=-ig\langle[P,\partial_{\mu}P]V^{\mu}\rangle,
\end{align}
the two body decay width $\Gamma(K^{*+}\to{}K^0\pi^{+})$ is related to $g$ as
\begin{align}
\Gamma(K^{*+}\to{}K^0\pi^{+})=\frac{g^2}{6\pi{}m^2_{K^{*+}}}{\cal{P}}^3_{\pi{}K^{*}}=\frac{2}{3}\Gamma_{K^{*+}},
\end{align}
where the ${\cal{P}}_{\pi{}K^{*}}$ is the three momentum of the $\pi$ in the rest frame of the $K^{*}$.  Using the experimental strong
decay width $\Gamma_{K^{*+}}=50.3\pm 0.8$ MeV and the masses of the particles~\cite{ParticleDataGroup:2022pth}, we obtain $g=4.64$.

\begin{table*}[t]
	\centering \tabcolsep=1mm \renewcommand\arraystretch{1.5}
	\fontsize{8}{11}\selectfont
	\caption{The coefficients for the $t$-channel mesons exchange potential. }\label{table81a}
	\begin{tabular}{c|ccc|ccc|cc|cc|c|cc|ccccccccccccccccc}
		\hline\hline
		\multirow{2}{*}{channel (j)}  &\multicolumn{3}{c|}{${\cal{I}}_V$} &\multicolumn{3}{c|}{${\cal{J}}_V$ } &\multicolumn{2}{c|}{${\cal{Z}}_P$ } &\multicolumn{2}{c|}{${\cal{S}}_P$}
        &\multirow{2}{*}{channel (j)} &${\cal{I}}_V$                      &${\cal{J}}_V$     &  ${\cal{Z}}_P$  &  ${\cal{S}}_P$\\
		                                                   \cline{2-4}                         \cline{5-7}         \cline{8-9}   \cline{10-11}    \cline{13-16}
		  &$\rho$ & $\omega$ &$\phi$   &$\rho$ & $\omega$ &$\phi$  & $\eta$  &$\pi$  & $\eta$  &$\pi$
          &       & $\bar{K}^{*}$      &$\bar{K}^{*}$  &  \multicolumn{2}{c}{$\bar{K}$ } \cr  \toprule \hline
		
		%\cmidrule(lr){1-5}
		%\multirow{5}{*}
		$K^{*-}\Sigma^{+}\to{}K^{*-}\Sigma^{+}$           &    1       &1         &-1   &$F$                  &$D$                  & 0              &$\frac{D_1}{6}$  &$\frac{F_1}{2}$
        &$-\frac{D_1}{3}$       & 0  &$K^{*-}\Sigma^{+}\to\rho^{+}\Xi^{-}$  &0 &0 &0 &0\\
		$\bar{K}^{*0}\Sigma^{0}\to{}\bar{K}^{*0}\Sigma^0$ &    0       &1         &-1   &0                    &$D$                  & 0              &$\frac{D_1}{6}$  & 0
        &$-\frac{D_1}{3}$       & 0  &$K^{*-}\Sigma^{+}\to\rho^{0}\Xi^{0}$  &$-\frac{1}{\sqrt{2}}$ &$-\frac{F+D}{\sqrt{2}}$ &0 &$\frac{D_1}{2\sqrt{2}}$ \\
		$K^{*-}\Sigma^{+}\to{}\bar{K}^{*0}\Sigma^{0}$     &$-\sqrt{2}$ &0         & 0   &$-\sqrt{2}F$         & 0                   & 0              &0        &$-\frac{F_1}{\sqrt{2}}$
        &0                      & 0 &$\bar{K}^{*0}\Sigma^{0}\to\rho^{0}\Xi^{0}$  &$-\frac{1}{2}$ &$-\frac{F+D}{2}$ &0 &$\frac{F_1}{4}$ \\
        $\bar{K}^{*0}\Lambda\to{}\bar{K}^{*0}\Lambda$     &    0       &$\sqrt{2}$&$-1$ &0                    &$\frac{\sqrt{2}D}{3}$& $-\frac{4D}{3}$&$-\frac{D_1}{6}$ & 0
        &$\frac{D_1}{3}$        & 0 &$\bar{K}^{*0}\Sigma^{0}\to\rho^{+}\Xi^{-}$  &$-\frac{1}{2}$ &$-\frac{F+D}{2}$ &0 &$\frac{D_1}{2\sqrt{2}}$ \\
         $\rho^{+}\Xi^{-}\to{}\rho^{+}\Xi^{-}$            &   -1       &0         & 0   &$D-F$                & 0                   & 0              &$-\frac{D_1}{12}$&$-\frac{F_1}{4}$  &$-\frac{D_1}{12}$      &$\frac{F_1}{4}$ &$\bar{K}^{*0}\Lambda\to\rho^{+}\Xi^{-}$ &$-\frac{\sqrt{6}}{2}$ &$\frac{D-3F}{\sqrt{6}}$ &0 &$\frac{\sqrt{6}F_1}{4}$\\
         $\rho^{0}\Xi^{0}\to{}\rho^{0}\Xi^{0}$            &    0       &0         & 0   &0                    & 0                   & 0              &$-\frac{D_1}{12}$&0  &$-\frac{D_1}{12}$      &0  &$\bar{K}^{*0}\Lambda\to\rho^{+}\Xi^{-}$ &$-\frac{\sqrt{6}}{2}$ &$\frac{D-3F}{\sqrt{6}}$ &0 &$\frac{\sqrt{6}F_1}{4}$\\
         $\rho^{+}\Xi^{-}\to{}\rho^{0}\Xi^{0}$            &$-\sqrt{2}$ &0         & 0   &$\sqrt{2}(D-F)$      & 0                   & 0              &0        &$\frac{D_1}{2\sqrt{2}}$   &0        &$-\frac{D_1}{2\sqrt{2}}$ &$K^{*-}\Sigma^{+}\to\omega\Xi^{0}$ &$-\frac{1}{\sqrt{2}}$ &$-\frac{D+F}{\sqrt{2}}$ &0 &$\frac{F_1}{2\sqrt{2}}$\\
         $\omega\Xi^{0}\to{}\omega\Xi^{0}$                &    0       &0         & 0   &0                    & 0                   & 0              &$-\frac{F_1}{4}$ &0&0&0
         &$\bar{K}^{*0}\Sigma^{0}\to\omega\Xi^{0}$ &$\frac{1}{2}$ &$\frac{D+F}{2}$ &0 &$-\frac{D_1}{4}$\\
         $\phi\Xi^{0}\to{}\phi\Xi^{0}$                    &    0       &0         & 0   &0                    & 0                   & 0              &$\frac{F_1}{2}$ &0&0&0
         &$\bar{K}^{*0}\Lambda\to\omega\Xi^{0}$ &$-\frac{\sqrt{3}}{2}$ &$\frac{D-3F}{2\sqrt{3}}$ &0 &$-\frac{D_1}{4\sqrt{3}}$\\
         $K^{*-}\Sigma^{+}\to{}\bar{K}^{*0}\Lambda$       &    0       &0         & 0   &$\frac{\sqrt{6}D}{3}$& 0                   & 0              &0   &$\frac{D_1}{\sqrt{6}}$ &0 &0   &$K^{*-}\Sigma^{+}\to\phi\Xi^{0}$ &1  &1 &$\frac{F_1}{2}$ &0\\
         $\bar{K}^{*0}\Sigma^{0}\to{}\bar{K}^{*0}\Lambda$ &    0       &0         & 0   &$-\frac{D}{\sqrt{3}}$& 0                   & 0              &0   &$-\frac{D_1}{2\sqrt{3}}$ &0 &0    &$\bar{K}^{*0}\Sigma^{0}\to\phi\Xi^{0}$ &$-\frac{1}{\sqrt{2}}$  &$-\frac{D+F}{\sqrt{2}}$ &$-\frac{F_1}{2\sqrt{2}}$ &0\\
         $\rho^{+}\Xi^{-}\to\omega\Xi^0$                  &    0       &0         & 0   &0                    & 0                   & 0              &0   &$-\frac{F_1}{2\sqrt{2}}$ &0 &$-\frac{F_1}{2\sqrt{2}}$   &$\bar{K}^{*0}\Lambda\to\phi\Xi^{0}$ &$\frac{\sqrt{6}}{2}$  &$\frac{3F-D}{\sqrt{6}}$ &$-\frac{D_1}{2\sqrt{6}}$ &0\\
         $\rho^{0}\Xi^{0}\to\omega\Xi^0$                  &    0       &0         & 0   &0                    & 0                   & 0              &0   &$-\frac{D_1}{4}$ &0 &$-\frac{D_1}{4}$   \\
         $\rho^{+}\Xi^{-}\to\phi\Xi^0$                    &    0       &0         & 0   &0                    & 0                   & 0              &0   &0   &0   &0  \\
         $\rho^{0}\Xi^{0}\to\phi\Xi^0$                    &    0       &0         & 0   &0                    & 0                   & 0              &0   &0   &0   &0   \\
          $\omega\Xi^{0}\to\phi\Xi^0$                     &    0       &0         & 0   &0                    & 0                   & 0              &0   &0   &0   &0   \\
		\hline\hline
	\end{tabular}\vspace{0cm}
\end{table*}
\begin{table}[h!]
	\centering \tabcolsep=1mm \renewcommand\arraystretch{1.5}
	\fontsize{8}{11}\selectfont
	\caption{${\cal{C}}_{j_k}$ is the isospin coefficient with $k=1,2$.
     }\label{table812}
	\begin{tabular}{cc|ccc|ccc|ccc|ccc|ccc}
		\hline\hline
     \multicolumn{2}{c|}{\multirow{2}*{$j_{k}$}}  &\multicolumn{3}{c|}{$\Sigma\bar{K}^{*}$} &\multicolumn{3}{c|}{$\Lambda\bar{K}^{*}$ } &\multicolumn{3}{c|}{$\Xi\rho$ }
        &\multicolumn{3}{c|}{$\Xi\omega$}   &\multicolumn{3}{c}{$\Xi\phi$}  \\
		   \cline{3-5}         \cline{6-8}             \cline{9-11}        \cline{12-14}        \cline{15-17}
        \multicolumn{2}{c|}{} &$\Sigma^{+}K^{*-}$& &$\Sigma^{0}\bar{K}^{*0}$& &$\Lambda\bar{K}^{*0}$ & &$\Xi^{-}\rho^{+}$&  & $\Xi^{0}\rho^{0}$ && $\Xi^{0}\omega$ & &&$\Xi^{0}\phi$&\cr\toprule \hline
		
		%\cmidrule(lr){1-5}
		%\multirow{5}{*}
     \multicolumn{2}{c|}{${\cal{C}}_{j_k}$}  &$\sqrt{\frac{2}{3}}$ & &$-\frac{1}{\sqrt{3}}$  &  &$1$ & &$-\sqrt{\frac{2}{3}}$ &&$\frac{1}{\sqrt{3}}$   &&$1$&  &&$1$ & \\
		\hline\hline
	\end{tabular}\vspace{0cm}
\end{table}

To compute the potential corresponding to Fig.~\ref{fig.exchange}(a), the chiral Lagrangian for meson-baryon interactions are
needed~\cite{Garzon:2012np}
\begin{align}
{\cal{L}}_{PBB}=&-\frac{\sqrt{2}}{f}\frac{D_1+F_1}{2}\langle\bar{B}\gamma_{\mu}\gamma_5\partial^{\mu}PB\rangle\nonumber\\
                &-\frac{\sqrt{2}}{f}\frac{D_1-F_1}{2}\langle\bar{B}\gamma_{\mu}\gamma_5B\partial^{\mu}P\rangle,
\end{align}
where $F_1=0.51$, $D_1=0.75$, and $f=93$ MeV~\cite{Garzon:2012np}.

Because hadrons are not pointlike particles, we need to include the form factors in evaluating the scattering
amplitudes.  For the $t$-channel mesons exchange, we would like to apply a widely used pole form factor, which is
\begin{align}
{\cal{F}}_{i}=\frac{\Lambda^2-m_i^2}{\Lambda^2-q^2}=\frac{\Lambda^2-m_i^2}{\tilde{\Lambda}^2+q^2},
\end{align}
where $m_i$ and $q$ are the masses and the four-momenta of exchanged mesons, respectively.  The cutoff $\Lambda=m_i+\alpha{}\Lambda_{QCD}$
with $\Lambda_{QCD}$=220 MeV, and $\tilde{\Lambda}=\Lambda^2-(m_f^V-m_i^V)^2$.  The parameter $\alpha$ is taken as a parameter and discussed later.

Putting all the pieces together,  the effective potentials can be easily computed.  For the $t$-channel mesons exchange, the potential are written as
\begin{align}
{\cal{V}}^{t}_{j}&=\sum_{j}\prod_{k=1,2}{\cal{C}}_{j_k}(\sum_{{\cal{H}}=V,P}{\cal{M}}^{{\cal{H}}}_{j}),
\end{align}
with
\begin{align}
{\cal{M}}^{V}_j&=-ig\bar{u}_{\lambda_1^{'}}(q_1)[{\cal{X}}_Vg\gamma_{\eta}-\frac{{\cal{Y}}_V}{4M_{\Sigma}}(\gamma_{\eta}q\!\!\!/_{t}-q\!\!\!/_{t}\gamma_{\eta})]\nonumber\\
                 &\times{}u_{\lambda_1}(p_1)\frac{(-g^{\eta\sigma}+q^{\eta}_tq^{\sigma}_t/m^2_V)}{q^2_t-m^2_{V}}[(q_t-p_2)\cdot\epsilon^{\dagger}\nonumber\\
                 &\epsilon_{\sigma}-(q_t+q_2)\cdot\epsilon\epsilon_{\sigma}^{\dagger}+(p_2+q_2)_{\sigma}\epsilon^{\dagger}\cdot\epsilon],
\end{align}
\begin{align}
{\cal{M}}^{P}_{j}&=\frac{G^{'}}{f}\bar{u}_{\lambda_1^{'}}(q_1)q\!\!\!/_t\gamma_5u_{\lambda_1}(p_1)\frac{1}{q_t^2-m^2_{P}}\epsilon_{\mu\nu\alpha\beta}\nonumber\\
                 &\times({\cal{Z}}_Pq_2^{\mu}\epsilon^{\dagger}_{\nu}p_2^{\alpha}\epsilon_{\beta}+{\cal{S}}_Pp_2^{\mu}\epsilon_{\nu}q_2^{\alpha}\epsilon^{\dagger}_{\beta}),
\end{align}
where $u$ is the Dirac spinor of baryon, and $\epsilon_{\sigma}$ is the polarization vector of meson.  $q_t=q_2-p_2$ is the four momenta of the $t$-channel
exchange mesons, and $q\!\!\!/_t=q_t^{\mu}\gamma_{\mu}$.   The index $j$ stand by the different channels, and $j_1$ and $j_2$ is its initial state and final
state, respectively.   The coefficients ${\cal{X}}_V$, ${\cal{Y}}_V$ for vector mesons exchange potential and ${\cal{Z}}_P$, ${\cal{S}}_P$ for pseudoscalar meson
exchange potentials are computed in Tab.~\ref{table81a}.  ${\cal{C}}_{j_k}$ is the isospin coefficient, which are calculated from the following isospin assignments~\cite{Oset:1997it}
\begin{align}
&\left(
  \begin{array}{cc}
    p  \\
    n \\
  \end{array}
\right)=\left(
  \begin{array}{cc}
   |\frac{1}{2},\frac{1}{2}\rangle \\
   |\frac{1}{2},-\frac{1}{2}\rangle \\
  \end{array}
\right),~~~~~
\left(
  \begin{array}{cc}
    K^{*-} \\
   \bar{K}^{*0} \\
  \end{array}
\right)=\left(
  \begin{array}{cc}
   -|\frac{1}{2},-\frac{1}{2}\rangle \\
    |\frac{1}{2},\frac{1}{2}\rangle \\
  \end{array}
\right),\nonumber\\
&\left(
  \begin{array}{cc}
    \rho^{+}   \\
    \rho^0     \\
    \rho^{-}   \\
  \end{array}
\right)=\left(
  \begin{array}{cc}
   |1,1\rangle \\
   |1,0\rangle \\
   |1,-1\rangle \\
  \end{array}
\right),~~~~~
\left(
  \begin{array}{cc}
    \Sigma^{+} \\
    \Sigma^0   \\
    \Sigma^{-} \\
  \end{array}
\right)=\left(
  \begin{array}{cc}
   -|1,1\rangle  \\
    |1,0\rangle  \\
    |1,-1\rangle \\
  \end{array}
\right),\nonumber\\
&\left(
  \begin{array}{cc}
   \Xi^{0} \\
   \Xi^{-} \\
  \end{array}
\right)=\left(
  \begin{array}{cc}
   |\frac{1}{2},\frac{1}{2}\rangle  \\
  -|\frac{1}{2},-\frac{1}{2}\rangle \\
  \end{array}
\right),\nonumber
\end{align}
and we collect them in Tab.~\ref{table812}.

In the above equation, the polarization vector $\epsilon^{\mu}(s)$ represent the wave function of
the spin-1 field and can be expressed as
\begin{align}
\epsilon^{\mu}(s=0)
  =\left(
  \begin{array}{cc}
    0   \\
    0   \\
    0   \\
    1   \\
  \end{array}
\right);~~~~
\epsilon^{\mu}(s=\pm)
  =\frac{1}{\sqrt{2}}
   \left(
  \begin{array}{cc}
    0      \\
    \mp1   \\
    -i   \\
    0   \\
  \end{array}
\right).
\end{align}
The $u_{s_1}(p_1)$ and $\bar{u}_{s^{'}_1}(q_1)$ stand for the spin wave function of baryons.  In this work we adopt the Dirac spinor as
\begin{align}
u(\vec{q},s)&=\sqrt{\frac{E+m}{2m}}
\left(
  \begin{array}{ccc}
    1\\
    \frac{\vec{\sigma}\cdot\vec{q}}{E+m}\\
  \end{array}
\right){\cal{\chi}}_{s},\\
\bar{u}(\vec{q},s)&=u^{\dagger}(\vec{q},s)\gamma^0={\cal{\chi}}^{\dagger}_{s}\sqrt{\frac{E+m}{2m}}
\left(
  \begin{array}{ccc}
    1 & -\frac{\vec{\sigma}\cdot\vec{q}}{E+m}\\
  \end{array}
\right),
\end{align}
where ${\cal{\chi}}_{1/2}=(1,0)^{\dagger}$ and ${\cal{\chi}}_{-1/2}=(0,1)^{\dagger}$, and index $\pm{1/2}$ are the third component of the spin $s$.  $E$, $\vec{q}$, and $m$
are the energy, three-momenta, and masses of baryons, respectively, and they have on-shell relation $E^2=\vec{q}^2+m^2$.

By considering the nonrealistic approximation and keep the terms up to order of $1/m^2$, the scattering amplitudes are
\begin{align}
{\cal{M}}^{P}_{j}&=\frac{G^{'}}{f}\frac{\vec{\sigma}\cdot\vec{q}}{\vec{q}^2+\omega^2_P}({\cal{Z}}_P+{\cal{S}}_P)[(m^{V}_{f}-m_i^{V})\vec{k}\cdot{}(\vec{\epsilon}\times\vec{\epsilon}^{\dagger})\nonumber\\
                 &-\frac{(m^{V}_f+m^{V}_i)}{2}\vec{q}\cdot{}(\vec{\epsilon}\times\vec{\epsilon}^{\dagger})],\\
{\cal{M}}^{V}_j&=\frac{ig}{\vec{q}^2+\omega^2_{V}}(-g{\cal{X}}_V{\cal{T}}_1+\frac{{\cal{Y}}_V}{4m_i^B}{\cal{T}}_2),
\end{align}
with
\begin{align}
\omega_P^2&=m_P^2-(m_f^V-m_i^V)^2,\\
\omega_V^2&=m_V^2-(m_f^V-m_i^V)^2,\\
{\cal{T}}_1&=\frac{m_f^{V}-m_i^{V}}{m_V^2}\vec{A}\cdot{}\vec{q}+(\frac{(m_i^B+m_f^B)}{2m_i^Bm_f^B}\vec{\sigma}_2\cdot\vec{k}+\frac{(m_i^B-m_f^B)}{4m_i^Bm_f^B}\nonumber\\
           &\times\vec{\sigma}_2\cdot\vec{q})[\vec{\sigma}_{1}\cdot{}\vec{A}-\frac{(m_f^{V2}-m_i^{V2})}{m_V^2}(\vec{\sigma}_{1}\cdot{}\vec{q})\epsilon^{\dagger}\cdot\epsilon]\nonumber
\end{align}
\begin{align}
&+(1+\frac{\vec{k}^2-\frac{1}{4}\vec{q}^2}{4m_f^Bm_i^B})[1-\frac{(m_f^{V}-m_i^{V})^2}{m_V^2}](m_f^V+m_i^V)\epsilon^{\dagger}\cdot\epsilon,\\
{\cal{T}}_2&=[(m_f^{V}-m_i^{V})(\vec{\sigma}_{1}\cdot{}\vec{A})-(m_f^V+m_i^V)\epsilon^{\dagger}\cdot\epsilon(\vec{\sigma}_{1}\cdot{}\vec{q})]\nonumber\\
           &\times(\frac{(m_i^B-m_f^B)}{m_i^Bm_f^B}\vec{\sigma}_2\cdot\vec{k}+\frac{(m_i^B+m_f^B)}{2m_i^Bm_f^B}\vec{\sigma}_2\cdot\vec{q}),
\end{align}
where $\vec{A}=(\frac{3}{2}\vec{q}-\vec{k})\cdot\vec{\epsilon}^{\dagger}\vec{\epsilon}-(\frac{3}{2}\vec{q}+\vec{k})\cdot\vec{\epsilon}\vec{\epsilon}^{\dagger}+2\vec{k}\epsilon^{\dagger}\cdot\epsilon$,
and $m_i$ and $m_f$ are the masses of the initial states and final states, respectively.   Variables in the above functions denote $\vec{q}=\vec{q}_2-\vec{p}_2$ and $\vec{k}=\frac{1}{2}(\vec{q}_2+\vec{p}_2)$.   Note that the $\omega_P$ and $\omega_V$ are always positive for considering systems.

\begin{table*}[t!]
	\centering
	\caption{The matrix elements of two-body interaction operators for $VB$ systems.}\label{tableqw}
	\begin{tabular}{cccccccc}
		\hline\hline
		$VB\to{}VB$    &$1/2^{-}$                    & $3/2^{-}$                               &$5/2^{-}$                   &$1/2^{+}$                   &$3/2^{+}$                               &$5/2^{+}$   \\
${\cal{Z}}$		&($^{2}S_{1/2}$,$^{4}D_{1/2}$)& ($^{4}S_{3/2},^{2}D_{3/2},^{4}D_{3/2}$) &($^{2}D_{5/2},^{4}D_{5/2}$) &($^{2}P_{1/2},^{4}P_{1/2}$) &($^{2}P_{3/2},^{4}P_{3/2},^{4}F_{3/2}$) &($^{4}P_{5/2},^{2}F_{5/2},^{4}F_{5/2}$)\\
	$\vec{\sigma}\cdot(-i\vec{\epsilon}_{s_3}^\dagger\times\vec{\epsilon}_{s_1})$&
		$\left(
		\begin{array}{cc}
			-2        ~~~&   0        \\
			0       ~~~&   1         \\
		\end{array}
		\right)$  &
		$\left(
		\begin{array}{ccc}
			1        &  0         &   0     \\
			0        &  -2        &   0     \\
			0        &  0         &   1     \\
		\end{array}
		\right)$ &
		$\left(
		\begin{array}{cc}
			-2        ~~~&   0        \\
			0       ~~~&   1         \\
		\end{array}
		\right)$&
		$\left(
		\begin{array}{cc}
			-2        ~~~&   0        \\
			0       ~~~&   1         \\
		\end{array}
		\right)$&
		$\left(
		\begin{array}{ccc}
			-2       &  0         &   0     \\
			0        &  1         &   0     \\
			0        &  0         &   1     \\
		\end{array}
		\right)$&
		$\left(
		\begin{array}{ccc}
			1        &  0         &   0     \\
			0        &  -2        &   0     \\
			0        &  0         &   1     \\
		\end{array}
		\right)$
		\\	
$S(\hat{r},\vec{\sigma},-i\vec{\epsilon}^{\dagger}_{s_3}\times{}\vec{\epsilon}_{s_1})$
		&$\left(\begin{array}{cc}
			0          &  -\sqrt{2}   \\
			-\sqrt{2}   &  -2          \\
		\end{array}\right) $
		& $\left(
		\begin{array}{ccc}
			0  &  1   &   2     \\
			1  &  0   &   -1    \\
			2  &  -1  &   0    \\
		\end{array}
		\right)$
		& $\left(
		\begin{array}{cc}
			0           &  \sqrt{\frac{2}{7}}   \\
			\sqrt{\frac{2}{7}}  &  \frac{10}{7} \\
		\end{array}
		\right)$
		&$\left(
		\begin{array}{cc}
			0          &  -\sqrt{2}   \\
			-\sqrt{2}  &  -2          \\
		\end{array}
		\right)$
		&$\left(
		\begin{array}{ccc}
			0                   &  \frac{1}{\sqrt{5}}  &   -\frac{3}{\sqrt{5}}    \\
			\frac{1}{\sqrt{5}}  &  \frac{8}{5}         &   \frac{6}{5}            \\
			-\frac{3}{\sqrt{5}} &  \frac{6}{5}         &   -\frac{8}{5}           \\
		\end{array}
		\right)$
		&$\left(
		\begin{array}{ccc}
			-\frac{2}{5}        &  \sqrt{\frac{6}{5}}        &   \frac{4\sqrt{6}}{5}     \\
		 \sqrt{\frac{6}{5}}         &  0                   &   -\frac{2}{\sqrt{5}}     \\
			\frac{4\sqrt{6}}{5}  &  -\frac{2}{\sqrt{5}} &   \frac{2}{5}             \\
		\end{array}
		\right)$ \\
$\vec{\epsilon}_{s_1}\cdot\vec{\epsilon}^{\dagger}_{s_3}$	
           &$\left(\begin{array}{cc}
				1        & 0   \\
			0  &  1          \\
			\end{array}\right) $
			& $\left(
			\begin{array}{ccc}
			1 &  0   &   0    \\
				0 &  1 &   0    \\
				0  &  0  &   1    \\
			\end{array}
			\right)$
			& $\left(
			\begin{array}{cc}
			1          & 0   \\
			0  & 1\\
			\end{array}
			\right)$
			&$\left(
			\begin{array}{cc}
				1        & 0  \\
				0  &  1          \\
			\end{array}
			\right)$
			&$\left(
			\begin{array}{ccc}
			1                   & 0  &   0 \\
			0  &  1         &  0           \\
				0 &  0        &  1    \\
			\end{array}
			\right)$
			&$\left(
			\begin{array}{ccc}
			1        & 0      &  1    \\
			0      &  1                &   0    \\
			0  & 0 &  1     \\
			\end{array}
			\right)$ \\
$S(\hat{r},\vec{\epsilon}_{s1},\vec{\epsilon}^{\dagger}_{s_3})$	
             &$\left(\begin{array}{cc}
				0          & -\sqrt{2}   \\
			    -\sqrt{2}  &  1          \\
			\end{array}\right) $
            & $\left(
			\begin{array}{ccc}
			0        &  1      &  -1    \\
			1        &  0      &  -1    \\
			-1       &  -1     &  0    \\
			\end{array}
			\right)$
           & $\left(
		\begin{array}{cc}
			0           &  \sqrt{\frac{2}{7}}   \\
			\sqrt{\frac{2}{7}}  &  -\frac{5}{7} \\
		\end{array}
		\right)$
        &$\left(\begin{array}{cc}
				0          & -\sqrt{2}   \\
			    -\sqrt{2}  &  1          \\
			\end{array}\right) $
        &$\left(
		\begin{array}{ccc}
			0                   &  \frac{1}{\sqrt{5}}  &   -\frac{3}{\sqrt{5}}    \\
			\frac{1}{\sqrt{5}}  &  -\frac{4}{5}        &   -\frac{3}{5}            \\
			-\frac{3}{\sqrt{5}} &  -\frac{3}{5}         &   \frac{4}{5}           \\
		\end{array}
		\right)$
        &$\left(
		\begin{array}{ccc}
			\frac{1}{5}        &  \sqrt{\frac{6}{5}}        &   -\frac{2\sqrt{6}}{5}     \\
		 \sqrt{\frac{6}{5}}         &  0                   &   -\frac{2}{\sqrt{5}}     \\
			-\frac{2\sqrt{6}}{5}  &  -\frac{2}{\sqrt{5}} &   -\frac{1}{5}             \\
		\end{array}
		\right)$\\
    $\vec{\sigma}\cdot\vec{L}\vec{\epsilon}_{s1}\cdot\vec{\epsilon}^{\dagger}_{s_3}$	
    & $\left(\begin{array}{cc}
				0        & 0   \\
			    0        & -3  \\
			\end{array}\right) $
    & $\left(
			\begin{array}{ccc}
			0        &  0      &  0    \\
			0        &  1      &  -2    \\
			0        &  -2     &  -2    \\
			\end{array}
			\right)$
    & $\left(\begin{array}{cc}
				-\frac{2}{3}             & -\frac{2\sqrt{14}}{3}   \\
			    -\frac{2\sqrt{14}}{3}    & -\frac{1}{3}  \\
			\end{array}\right) $
    & $\left(\begin{array}{cc}
				\frac{2}{3}             & -\frac{2\sqrt{2}}{3}   \\
			    -\frac{2\sqrt{2}}{3}    & -\frac{5}{3}  \\
			\end{array}\right) $
    &$\left(
		\begin{array}{ccc}
			-\frac{1}{3}          &  -\frac{2\sqrt{5}}{3}  &   0    \\
			-\frac{2\sqrt{5}}{3}  &  -\frac{2}{3}          &   0            \\
			       0              &  0                     &   -4           \\
		\end{array}
		\right)$
    &$\left(
		\begin{array}{ccc}
			1          &    0     &       0    \\
			0          &  \frac{4}{3}          &   -\frac{4\sqrt{5}}{3}            \\
			0          &  -\frac{4\sqrt{5}}{3}                     &   -\frac{7}{3}           \\
		\end{array}
		\right)$\\
		\hline
		\hline
	\end{tabular}
\end{table*}

Now we make the following Fourier transformation to get the effective potential $V(r)$ in the coordinate space,
\begin{align}
{\cal{V}}_j(r)=\int\frac{d^3\vec{q}}{(2\pi)^3}e^{-i\vec{q}\cdot{}\vec{r}}{\cal{V}}_j(\vec{q}){\cal{F}}^2(\vec{q}),
\end{align}
where the momentum-space potentials ${\cal{V}}(\vec{q})$ is obtained by utilizing Breit approximation~\cite{Breit:1929zz,Breit:1930zza}
\begin{align}
{\cal{V}}_j(\vec{q})=-\frac{\mathcal{M}_j^{t}}{\sqrt{\prod_{i}2{m}_{i}\prod_{f}2{m}_f}}.
\end{align}
Thus, the expressions of corresponding coordinate-space potentials can be obtained with the relation
\begin{align}
{\cal{F}}&\{\frac{1}{\vec{q}^2+a^2}(\frac{\Lambda^2-m^2}{b^2+\vec{q}^2})^2\}= Y_{1}(\Lambda,m,a,b,r),\nonumber\\
{\cal{F}}&\{\frac{\vec{q}^2}{\vec{q}^2+a^2}(\frac{\Lambda^2-m^2}{b^2+\vec{q}^2})^2\}=-\nabla_r^2Y_1(\Lambda,m,a,b,r),\nonumber\\
{\cal{F}}&\{\frac{\vec{k}^2}{\vec{q}^2+a^2}(\frac{\Lambda^2-m^2}{b^2+\vec{q}^2})^2\}=\frac{1}{4}\nabla_r^2Y_1(\Lambda,m,a,b,r)-\dfrac{{\cal{F}}_1}{2},\nonumber\\
{\cal{F}}&\{\frac{i\vec{\sigma}\cdot(\vec{q}\times\vec{k})}{\vec{q}^2+a^2}(\frac{\Lambda^2-m^2}{b^2+\vec{q}^2})^2\}
         =\vec{\sigma}\cdot\vec{L}(\frac{1}{r}\frac{\partial}{\partial r})Y_1(\Lambda,m,a,b,r)\label{eq18},\nonumber\\
{\cal{F}}&\{\frac{(\vec{A}\cdot\vec{q})(\vec{B}\cdot \vec{q})}{\vec{q}^2+a^2}(\frac{\Lambda^2-m^2}{b^2+\vec{q}^2})^2\}
         =\frac{1}{3}(\vec{A}\cdot\vec{B})\nonumber\\
         &\times(-\nabla_r^2Y_1(\Lambda,m,a,r))+\frac{1}{3}S(\hat{r},\vec{A},\vec{B})\nonumber\\
         &\times(-r\frac{\partial}{\partial r}\frac{1}{r}\frac{\partial}{\partial r}Y_1(\Lambda,m,a,b,r)),\nonumber\\
{\cal{F}}&\{\frac{(\vec{A}\cdot\vec{k})(\vec{B}\cdot \vec{q})}{\vec{q}^2+a^2}(\frac{\Lambda^2-m^2}{b^2+\vec{q}^2})^2\}
       =\frac{1}{3}(\vec{A}\cdot\vec{B})\nonumber\\
        &\times(\nabla^2_rY_1(\Lambda,m,a,b,r))-\frac{1}{3}(-r\frac{\partial}{\partial r}\frac{1}{r}\frac{\partial}{\partial r}Y_1(\Lambda,m,a,b,r))\nonumber\\
         &-\frac{2}{3}(\frac{\partial}{\partial r}Y_1(\Lambda,m,a,b,r))(S(\hat{r},\vec{A},\vec{B})+\vec{A}\cdot\vec{B})\frac{\partial}{\partial{}r},\nonumber\\
{\cal{F}}&\{\frac{\vec{k}\cdot\vec{q}}{\vec{q}^2+a^2}(\frac{\Lambda^2-m^2}{b^2+\vec{q}^2})^2\}=ir \nabla{}Y_{1}(\Lambda,m,a,b,r),\nonumber\\
{\cal{F}}&\{\frac{(\vec{A}\cdot\vec{k})(\vec{B}\cdot \vec{k})}{\vec{q}^2+a^2}(\frac{\Lambda^2-m^2}{b^2+\vec{q}^2})^2\}
         =-\frac{1}{4}(\vec{A}\cdot\vec{B})\nabla^2 Y_{1}(\Lambda,m,a,b,r)\nonumber\\
         &+\frac{1}{3}(S(\hat{r},\vec{A},\vec{B})+\vec{A}\cdot\vec{B})[(\frac{2}{r}-\frac{\partial}{\partial{r}})Y_1\frac{\partial}{\partial{r}}-Y_1\nabla^2],
\end{align}
where ${\cal{F}}$ denotes the Fourier transformation, and $\nabla^2_r=\frac{1}{r^2}\frac{\partial}{\partial{r}}r^2\frac{\partial}{\partial{r}}$.  $\vec{\sigma}\cdot\vec{L}$
is the spin-orbital operator, and $S(\hat{r},\vec{x},\vec{y})=3(\hat{r}\cdot\vec{x})(\hat{r}\cdot\vec{y})-\vec{x}\cdot\vec{y}$
is tensor operator. The function
\begin{align}
Y_{1}&(\Lambda,m,a,b,r)=(\frac{\Lambda^2-m^2}{b^2-a^2})^2\nonumber\\
                     &\times[\frac{1}{4\pi{}r}(e^{-ar}-e^{-br})-\frac{b^2-a^2}{8\pi{}b}e^{-br}],
\end{align}
and
\begin{align}
{\cal{F}}_1=\{\nabla_r^2,Y_1(\Lambda,m,a,r)\}&=\nabla_r^2Y_1(\Lambda,m,a,r) \nonumber\\
                                              &+Y_1(\Lambda,m,a,r)\nabla_r^2.
\end{align}

In order to obtain numerical results, it is necessary to determine the values of spin-spin interactions and tensor force operators.
To perform the calculations, we must provide the spin-orbital wave functions for the systems under discussion. In principle, the
spin-orbital wave function can be uniquely defined by the orbital angular momentum ($L$), spin angular momentum ($S$), and total
angular momentum ($J$). This wave function can be expressed as:
\begin{align}
|VB(^{2S+1}L_J)\rangle&=\sum^{m_S,m_L}_{m,m^{'}}{\cal{C}}_{1/2m,1m^{'}}^{S,m_s}{\cal{C}}_{Sm_S,Lm_L}^{J,M}\nonumber\\
                                 &\times\chi_{1/2m}\epsilon^{\mu}_{m^{'}}|Y_{L,m_L}\rangle,
\end{align}
Here, ${\cal{C}}_{ab,cd}^{e,f}$ represents the Clebsch-Gordan coefficient, and $|Y_{L,m_L}\rangle$ stands for the spherical harmonics
function. $M$, $m_S$, and $m_L$ are the third components of the total angular momentum ($J$), spin angular momentum ($S$), and orbital
angular momentum ($L$), respectively. With this spin-orbital wave function, we can compute the spin-spin interaction operator and the
tensor operator, and the results are compiled in Table~\ref{tableqw}.

\section*{RESULTS AND DISCUSSIONS}\label{Sec: results}
In this study, we study the possibility of forming a molecular structure through $VB$=$(\bar{K}^{*}\Sigma/\rho\Xi/\bar{K}^{*}\Lambda/\phi\Xi/\omega\Xi)$
interactions.  We employ the one-boson exchange model, which can be adapted to construct of $VB\to VB$ interaction potentials.  The corresponding scattering
Feynman diagrams for the $VB\to VB$ reaction can be observed in Fig.~\ref{fig.exchange}, where we consider exchanges involving vector and pseudoscalar mesons
in the $t$-channel.  Utilizing the obtained effective potential, we investigate the possible bound state by solving the non-relativistic
Schr\"{o}dinger equation with the help of the Gaussian Expansion Method (GEM)
\begin{align}
-\frac{1}{2\mu}[\nabla_r^2-\frac{L(L+1)}{r^2}]\psi(\vec{r})+V(r)\psi(\vec{r})=E\psi(\vec{r})\label{eq46},
\end{align}
where $\mu$ represents the reduced mass of the discussed systems, calculated as $\mu=m_{V}m_{B}/(m_V+m_{B})$. The specific case of $L=0$ corresponds
to the $S$-wave, a well-studied aspect in previous works. For a detailed procedure on solving the Schr\"{o}dinger equation under the GEM framework,
readers can refer to Refs.~\cite{Hiyama:2003cu}, which we omit discussing here.

Since vector mesons $V=(\bar{K}^{*}/\rho/\phi/\omega)$ and baryons $B=(\Sigma/\Lambda/\Xi)$ carries spin parities $1^{-}$ and $1/2^{+}$, respectively,
an $S$-wave bound state should have spin parity $1/2^{-}$ or $3/2^{-}$.  At present, the spin-parity of $\Xi(2030)$ have not been fully determined~\cite{ParticleDataGroup:2022pth}.  Only an early experimental analysis suggested that the $\Xi(2030)$ should carry a spin $J \geq 5/2$~\cite{Amsterdam-CERN-Nijmegen-Oxford:1977bvi}.
It requires that the $\Xi(2030)$ is at least a $P$-wave state if it is a $VB$ molecular state.  We would like to note that in our formalism, the
partial wave decomposition is done only on $J^P$, and for a spin-parity state, it may contain contributions from different waves (see Tab.~\ref{tableqw}).
Here we do not distinguish the explicit contributions from which wave they come.  Generally, the contributions with smaller $L$
is more important for a certain spin parity close to threshold. In this work, we will consider the spin parities where at least $P$ wave is involved.  Since a
system with $J \geq 7/2$ will be a $D$- ($G$-) wave state, only the isospin $1/2$ systems with $J \leq 5/2$ will be considered in this work. Given
such constraints, possible bound state produced from the $VB$ interaction are listed in Table.~\ref{table3a} with the variation of the cutoff $\Lambda$.
\begin{table}[h!]
\fontsize{7}{11}\selectfont
\centering
\tabcolsep=2mm
\caption{Possible bound states for $VB$ system with $\Lambda_i=m_i+\alpha 220$ MeV and different spin-parity assignments (in units of MeV).
$E$ is the eigenvalue. Notation $\times$ means no binding solutions.}\label{table3a}
	\begin{tabular}{cccc|cccc}\hline\hline
			State               ~&$\alpha$    ~~&$E$  ~~&     &  State &$\alpha$           &$E$        &              \\ \hline
$J^P=1/2^{-}$~&  3.5        ~~&-11.64         ~~&         &$J^P=1/2^{+}$   &$4.0$       &-15.85     &              \\
			~&   3.6       ~~&-35.24          ~~&         &                &$4.1$       &-37.35     &              \\
			~&  3.7      ~~&-78.88            ~~&        &                &$4.2$        &-61.55     &              \\
			~&  3.8       ~~&-155.55          ~~&         &                &$4.3$        &-91.14     &              \\ \hline
$J^P=3/2^{+}$~&  3.2     ~~&-9.32             ~~&          &$J^P=3/2^{-}$   &$2.0$       &$-8.21 $            &\\
			~& 3.3       ~~&-24.21            ~~&         &                &$2.1$    &$-32.53 $            &       \\
			~& 3.4       ~~&-37.18            ~~&        &                &$2.2$    &$-53.36 $            &        \\
			~& 3.5       ~~&-50.68            ~~&         &                &$2.3$    &$-117.98 $            &      \\ \hline
$J^P=5/2^{+}$  ~&  $3.8$   ~~& $-5.69 $       ~~&      &$J^P=5/2^{-}$   &$\times$       &$\times $              &       \\
			   ~&  $3.9$   ~~& $-10.67$       ~~&       &                &$\times$    &$\times $            &        \\
			   ~&  $4.0$   ~~& $-18.32$       ~~&       &                &$\times$    &$\times $            &        \\
			   ~&  $4.1$   ~~& $-23.35$       ~~&       &                &$\times$    &$\times $            &        \\
			\hline\hline
		\end{tabular}
\end{table}

From Table.~\ref{table3a}, we can find that a bound state in the $I(J^P)=1/2(1/2^{-})$ case is obtained.  Its binding
energy is 11.64 MeV when the cutoff $\alpha$ is taken as $\alpha=3.5$.  And the binding energies of the bound states
produced from the $VB$ interaction increase with the cutoff $\alpha$.  In other words, as the parameter $\alpha$ increases
from small to large, the bound state become more and more tightly bound.  Additionally, we find the presence of another
$S$-wave bound state with $I(J^P)=1/2(3/2^{-})$.  This state is relatively loosely bound compared to the bound state obtained
in the $I(J^P)=1/2(1/2^{-})$ channel.  This is mainly because the bound state appears in the $I(J^P)=1/2(3/2^{-})$ channel
with a smaller $\alpha$ than that of the bound state in the $I(J^P)=1/2(1/2^{-})$ channel, and a smaller $\alpha$ corresponds
to a loosely bound state.  Obviously, these differences originate from the spin-orbit force contribution (see the fifth and seventh
rows of Tab.~\ref{tableqw}).

The existence of a bound state with a smaller value of $\alpha$ indicates that meson exchange contributions
are most significant in the $J^P=3/2^{-}$ channel, making it easily detectable. This suggests that for the $J^P=3/2^{-}$ channel,
The potential is considerably more attractive than that of the $J^P=1/2^{-}$ case, aligning with the conclusions drawn from Hund's
rule.

It is worth noting that the existence of possible $S$-wave $VB$ bound states are also discussed in Ref.~\cite{Khemchandani:2016ftn,Oset:2010tof,Gamermann:2011mq}.
Especially in Ref.~\cite{Khemchandani:2016ftn}, a coupled-channel unitary approach was adopted, leading to the discovery of a dynamically generated meson-baryon
molecular state with $I(J^P)=1/2(3/2^{-})$.  This bound state is strongly suggested to be the experiment observed state $\Xi(2120)$~\cite{Khemchandani:2016ftn}.
However, experimental information on the $\Xi(2120)$ is limited, making its assignment to a particular state challenging.  In our study, two states with
$I(J^P)=1/2(1/2^{-})$ and $I(J^P)=1/2(3/2^{-})$ are generated from the $VB$ interactions. Determining which one can be considered as
the $\Xi(2120)$ requires further investigation, especially concerning its strong decay width.  This is mainly because the strong
decay width heavily depends on the internal components of hadron.   Here, we find that if $\Xi(2120)$ is a bound state with $I(J^P)=1/2(1/2^{-})$,
it contains the dominant $\bar{K}^{*}\Sigma$ component and smaller but non-negligible $\bar{K}^{*}\Lambda$, $\rho\Xi$ and $\phi\Xi$ components.
However, almost only the $\bar{K}^{*}\Sigma$ component contributes to $\Xi(2120)$ if it is a $VB$ bound state with $I(J^P)=1/2(3/2^{-})$.
These comparisons are illustrated in Tab.~\ref{table4q}.
\begin{table}
  \fontsize{7}{11}\selectfont
  \centering
  \tabcolsep=2mm
  \caption{Different bound state components change with the cutoff parameter $\alpha$.}\label{table4q}
  \begin{tabular}{ccccccc}\hline\hline
  	 $$ & $\alpha$ & $K^{*}\Sigma(\%)$ & $K^{*}\Lambda(\%)$ & $\Xi\rho(\%)$ & $\Xi\omega(\%)$ & $\Xi\phi(\%)$ \\ \hline
  	 $J^{P}=1/2^{-}$ & 3.5 & 97.60 & 0.49 & 0.36 & 0.00017 & 1.37 \\
  	 & 3.6 & 97.15 & 0.70 & 0.48 & 0.00020 & 1.68 \\
  	 & 3.7 & 96.83 & 0.66 & 0.48 & 0.00024 & 1.8 \\
  	 & 3.8 & 96.38 & 0.77 & 0.54 & 0.00028 & 2.05 \\
  	 & 3.9 & 95.87 & 0.89 & 0.62 & 0.00032 & 2.32 \\ \hline
	$J^{P}=3/2^{-}$ & 2.0 & 99.95 & 0.0075 & 0.0066 & 0.0183 & 0.0259 \\
	& 2.1 & 99.93 & 0.0096 & 0.0081 & 0.0227 & 0.0320 \\
	& 2.2 & 99.92 & 0.012 & 0.0099 & 0.0278 & 0.0390 \\
	& 2.3 & 99.90 & 0.015 & 0.012 & 0.034 & 0.0471 \\
	& 2.4 & 99.88 & 0.0184 & 0.0144 & 0.0408 & 0.0565 \\ \hline
	$J^{P}=1/2^{+}$ & 4.0 & 98.65 & 0.2537 & 0.1691 & 0.3957 & 0.5329 \\
	& 4.1 & 98.48 & 0.2899 & 0.1902 & 0.4451 & 0.5995 \\
	& 4.2 & 98.28 & 0.3304 & 0.2134 & 0.4995 & 0.6728 \\
	& 4.3 & 98.07 & 0.3755 & 0.2389 & 0.5593 & 0.7534 \\
	& 4.4 & 97.84 & 0.4257 & 0.2669 & 0.6249 & 0.8417 \\ \hline
	$J^{P}=3/2^{+}$ & 3.2 & 99.55 & 0.2435 & 0.0593 & 0.1408 & 0.1937 \\
	& 3.3 & 99.48 & 0.0910 & 0.0683 & 0.1623 & 0.2233 \\
	& 3.4 & 99.40 & 0.1065 & 0.0785 & 0.1863 & 0.2564 \\
	& 3.5 & 99.32 & 0.1241 & 0.0898 & 0.2132 & 0.2935 \\
	& 3.6 & 99.22 & 0.1442 & 0.1024 & 0.2431 & 0.3348 \\ \hline
	$J^{P}=5/2^{+}$ & 3.8 & 99.77 & 0.0469 & 0.0268 & 0.0805 & 0.0883 \\
	& 3.9 & 99.74 & 0.0535 & 0.0300 & 0.0904 & 0.0990 \\
	& 4.0 & 99.71 & 0.0608 & 0.0336 & 0.1012 & 0.1106 \\
	& 4.1 & 99.67 & 0.0688 & 0.0374 & 0.1131 & 0.1232 \\
	& 4.2 & 99.63 & 0.0777 & 0.0416 & 0.1260 & 0.1370 \\
	\hline \hline
  \end{tabular}
\end{table}

\begin{figure}[h!]
\centering
\includegraphics[bb=20 70 800 355, clip, scale=0.40]{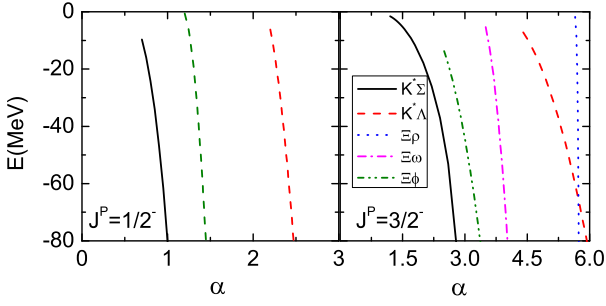}
\caption{The binding energy of bound states with different spin-parity obtained from the single-channel interactions as the function of the cutoff
$\alpha$.  The solid black line, dash red line, dot blue line, dash dot magenta line, and dash dot dot olive line represent the  bound state appear
in the $\bar{K}^{*}\Sigma$, $\bar{K}^{*}\Lambda$, $\rho\Xi$, $\omega\Xi$, and $\phi\Xi$ interactions, respectively. }\label{fig-ty.exchange}
\end{figure}

The individual contributions of the $\bar{K}^{*}\Sigma$, $\rho\Xi$, $\bar{K}^{*}\Lambda$, $\phi\Xi$, and $\omega\Xi$ channels for the bound states with
$I(J^P)=1/2(1/2^{-})$ and $I(J^P)=1/2(3/2^{-})$ are calculated and presented in Fig.~\ref{fig-ty.exchange}. It is found that the $\bar{K}^{*}\Sigma$,
$\phi\Xi$, and $\bar{K}^{*}\Lambda$ single-channel interactions play a dominant role, while single-channel interactions of the $\rho\Xi$ and $\omega\Xi$
has no contribution for the bound state in the $I(J^P)=1/2(1/2^{-})$ case.  However, for the bound state with $I(J^P)=1/2(3/2^{-})$, nearly all considered
single-channel interactions give contribution. By comparing with the results shown in Table \ref{table4q}, we can observe that interference interactions among
them are still significant.  For better clarity, let's use the bound state in the $I(J^P) = 1/2(1/2^{-})$ channel as an example.  We find that interference
interactions now lead to contributions from $\rho\Xi$ and $\omega\Xi$ to the bound state, whereas the pure $\rho\Xi$ and $\omega\Xi$ interactions made no
contribution.

By comparing, we also find that the size of molecular components is inversely proportional to the parameter $\alpha$. The smaller the $\alpha$,
the larger the ratio of the corresponding molecular component to the total molecular component.  We will also take the bound state in the
$I(J^P) = 1/2(1/2^{-})$ channel as an example to explain.  You can find from Fig.\ref{fig-ty.exchange} that the interaction contributions
correspond to the smallest $\alpha$ for the $\bar{K}^{*}\Sigma$ channel, intermediate $\alpha$ for the $\omega\Xi$ channel, and the largest
$\alpha$ for the $\bar{K}^{*}\Lambda$. However, you can clearly see that the primary component of the molecule is $\bar{K}^{*}\Sigma$, followed
by $\omega\Xi$, with $\bar{K}^{*}\Lambda$ being the smallest (see Tab.\ref{table4q}).   Similar conclusions can be obtained for other channels.
Moreover, the closer parameter $\alpha$ is to 1, the more it makes the corresponding component become the dominant component of the molecule.
This is likely why people generally consider the results obtained with $\alpha\simeq{}1$ in the one-boson exchange model to be the most reliable.
Nevertheless, based on experimental information, people obtained a relatively wide range of variations for the parameter $\alpha$~\cite{Dong:2017rmg,Chen:2012nva,Liu:2006df,Colangelo:2002mj,Meng:2007cx,Liu:2008tv,Xu:2015qqa}, which encompasses
the values considered in this work.

Now, we turn to discuss the possible high-wave molecular states related to the $\Xi(2030)$ through the $VB$ interactions, which have not been
previously explored in other works, such as those referenced in Refs.~\cite{Khemchandani:2016ftn,Oset:2010tof,Gamermann:2011mq}.  The experimental analysis~\cite{Amsterdam-CERN-Nijmegen-Oxford:1977bvi} favors a spin not less than $J=5/2$ for the $\Xi(2030)$, which is also suggested by
PDG~\cite{ParticleDataGroup:2022pth}.  As discussed above, the system with a spin parity $J^{P}=5/2^{-}$ and spin $J \geq 7/2$ is at least a
$D$-wave state, whose contribution should be small. Indeed, we find that the bound state with $I(J^P)=1/2(5/2^{-})$ is not survived (see Tab.~\ref{tableqw}).
That means the $\Xi(2030)$ state cannot be accommodated in the current $D$-wave $VB$ molecular picture.  The only possible spin-parity to interpret
the $\Xi(2030)$ in our molecular picture is $J^P=5/2^{+}$, corresponding to the $P/F$-wave $VB$ molecule (see Tab.~\ref{tableqw}).  Fortunately, a bound
state with $J^P=5/2^{+}$ can be found at a cutoff of about $\alpha=3.8$ in our model, as expected, and with the increase of the cutoff, its mass can reach
2030 MeV.  Notably, the study in Ref.~\cite{He:2016pfa} suggested that $P$-wave contributions are still considerable and may be observed in experiments.

In Table~\ref{table4q}, it is evident that the dominant contribution to the bound state with $J^P=5/2^{+}$ primarily arises from the $\bar{K}^{*}\Sigma$ component,
accounting for nearly 99.63\% to 99.77\% of the total component when $\alpha$ falls within the range of $\alpha=3.8-4.2$.  It is important to note that the repulsive
centrifugal force for the angular momentum ($L$) is the same in states with $J^P=5/2^{+}$ and $J^P=5/2^{-}$, as given by $L(L+1)/2\mu{}r^2$. However, we observe a
bound state in the case of $J^P=5/2^{+}$, while molecular formation is prohibited for $J^P=5/2^{-}$.  One possible explanation for this difference lies in the
significant variation of meson-exchange forces between these two cases.

In addition to the molecule discussed earlier, we are also interested in exploring other possible bound states arising from the $VB$ interaction.
Tab.~\ref{table3a} reveals the presence of a bound state with $J^P=1/2^{+}$ occurring around $\alpha=4.0-4.3$ and another bound state with
$J^P=3/2^{+}$ forming at $\alpha=3.2-3.5$. Given the predominant contribution of low partial waves, it is expected that these two molecular states
possess significant $P$-wave $VB$ components (refer to Tab.~\ref{tableqw}).

\section{Summary}\label{sec:summary}
Inspired by the LHCb observation of the pentaquark states and their molecular interpretations, we have studied possible bound states
from the $VB$=$(\bar{K}^{*}\Sigma/\rho\Xi/\bar{K}^{*}\Lambda/\phi\Xi/\omega\Xi)$ interaction by solving a Schr\"{o}dinger equation within
the one-boson exchange model.  A bound state with the quantum number $I(J^P)=1/2(5/2^+)$ from the $VB$ interaction is produced at $\alpha=3.8$.
This bound state can be associated to the $\Xi(2030)$ as a $P$-wave molecular state.  Four other bound states with quantum numbers $J^P = 1/2(1/2^{-})$,
$1/2(3/2^{-})$,and $1/2(1/2^{+}$, and $1/2(3/2^{+})$  are also produced from the $VB$=$(\bar{K}^{*}\Sigma/\rho\Xi/\bar{K}^{*}\Lambda/\phi\Xi/\omega\Xi)$
interaction.

For the $\Xi(2030)$, our study showed that it could be a $P$-wave $VB$ bound state with spin parity $J^P=5/2^{+}$, consistent with
the available experimental information.  On the other hand, we do not have enough experimental information to determine which bound
state produced from the $VB$ interaction is the $\Xi(2120)$.  If we follow the assignment in Ref.~\cite{Khemchandani:2016ftn}, that
the $\Xi(2120)$ is a molecular bound state with $J^P=1/2(3/2^{-})$, it is interesting to see that the $\Xi(2120)$ and $\Xi(2030)$
exhibit a pattern.  It is worth noting that if  $\Xi(2012)$ is an $S$-wave molecular state with $J^P=3/2^{-}$, we suggest
determining its spin and parity by studying its decay width, owing to the larger difference in their molecular components.

Clearly, more experimental efforts are needed to better understand the nature of the $\Xi(2120)$ and test the scenario proposed in
the present study. There exist plans to study double-strangeness baryons at facilities such as JLab, JPARC, and PANDA. We strongly
recommend our experimental colleagues to study the $\Xi(2030)$, $\Xi(2120)$, and the another double strangeness baryons.

\section*{Acknowledgments}
This work was supported by the National Natural Science Foundation
of China under Grant No.12005177.

\end{document}